\documentclass[aps,prl,twocolumn,superscriptaddress,amsmath,amssymb,citeautoscript]{revtex4}
\usepackage[utf8]{inputenc}
\usepackage[T1]{fontenc}
\usepackage{graphicx}
\usepackage{bm}
\usepackage{hyperref}
\usepackage[normalem]{ulem}

\begin{document}

\title{Phase coherence of charge-$6e$ superconductors via a frustrated Kagome XY antiferromagnet}
\author{Feng-Feng Song}
\affiliation{Institute for Solid State Physics, The University of Tokyo, Kashiwa, Chiba 277-8581, Japan}
\author{Guang-Ming Zhang}
\email{gmzhang@tsinghua.edu.cn}
\affiliation{State Key Laboratory of Low Dimensional Quantum Physics and Department of
Physics, Tsinghua University, Beijing 100084, China}
\affiliation{School of Physical Science and Technology, ShanghaiTech University, Shanghai,201210, China}
\affiliation{Frontier Science Center for Quantum Information, Beijing 100084, China}

\date{\today}

\begin{abstract}
Recent experimental evidence for the charge-$6e$ condensed phase in kagome superconductors has generated significant interest. We investigate the unconventional superconductivity in the kagome superconductor $\mathrm{CsV_3Sb_5}$, focusing on the emergence of charge-$6e$ superconductivity (SC) at temperatures higher than the conventional charge-$2e$ SC state. By modeling the phase coherence of the SC order parameter using a frustrated antiferromagnetic XY model on an emergent kagome lattice, we show that the condensation of fractional vortices with $1/3$ vorticity stabilizes phase coherence in $\exp(i3\theta)$, giving rise to the charge-$6e$ SC state. Using a tensor network approach tailored for frustrated spin systems, we identify a Berezinskii-Kosterlitz-Thouless transition at $T_c/J \simeq 0.075$, where the unbinding of $1/3$ fractional vortex-antivortex pairs transforms the system from the charge-$6e$ SC phase to the normal phase. Below $T_c$, the $1/3$ fractional vortex correlations exhibit power-law decay, while the integer vortex correlations decay exponentially, reflecting the dominance of charge-$6e$ SC in the absence of charge-$2e$ SC. Our results provide a theoretical understanding of the charge-$6e$ SC in two-dimensional kagome superconductors, emphasizing the interplay between fractional vortices, frustration, and topology in stabilizing this exotic SC phase.
\end{abstract}

\maketitle

{\textit{Introduction}}---The kagome lattice, with its frustrated geometry of corner-sharing triangles, has emerged as a versatile platform for studying the interplay between nontrivial band topology, strong correlations, and intertwined electronic orders. Its distinctive features, including flat bands, van Hove singularities, and the potential to host quantum spin liquids, have been widely explored~\cite{Balents_2010, Ortiz_2019, Broholm_2020, Neupert_2022, Jiang_2023}. Among kagome-based materials, the quasi-two-dimensional (2D) superconductors $\mathrm{AV_3Sb_5}$ (A = K, Rb, Cs) have garnered significant attention due to their unconventional charge density wave (CDW) states above subsequent superconductivity (SC) at low temperatures~\cite{Jiang_2021, Liang_2021, Li_2021, Ortiz_2021, Luo_2022, Hu_2022, Kang_2022, Nakayama_2022}. The CDW state in $\mathrm{AV_3Sb_5}$ exhibits unconventional symmetry breaking properties, including spontaneous time-reversal symmetry breaking (TRSB)~\cite{Jiang_2021, Mielke_2022, Le_2024} and rotation symmetry breaking~\cite{Zhao_2021, Chen_2021}, suggesting a complex intertwining of electronic orders.

Recently, $\mathrm{CsV_3Sb_5}$ has emerged as a promising platform to study unconventional SC. Magneto-transport experiments have revealed an exotic SC phase diagram, including evidence of charge-$6e$ superconductivity in the strongly fluctuating region above the conventional charge-$2e$ condensate~\cite{Ge_2024}. A hallmark of charge-$6e$ SC is the fractional magnetic flux quantization of $\frac{h}{6e}$, distinguishing it from the usual charge-$2e$ SC described by the Bardeen-Cooper-Schrieffer (BCS) theory. Charge-$6e$ SC is proposed to arise as a vestigial higher-order condensation of bound states of electron sextets, possibly driven by strong fluctuations of pair density wave (PDW) order~\cite{Agterberg_2008, Agterberg_2011, Agterberg_2020, Zhou_2022, Lin_2024}. In recent experiments, a commensurate $3\bm{Q}$ PDW order, where the SC order parameter spatially modulates along three distinct directions, has been observed in the SC state of $\mathrm{CsV_3Sb_5}$~\cite{Chen_2021, Han_2024, Deng_2024}, suggesting an intricate intertwining of SC and PDW as a potential mechanism for the emergence of charge-$6e$ SC.

Despite significant theoretical progress in recent works~\cite{Agterberg_2011, Zhou_2022}, the physical origin and implications of the remarkable observations of charge-$6e$ SC remain incompletely understood. To uncover the unconventional aspects of charge-$6e$ SC and the role of the underlying frustrated kagome geometry, it is crucial to understand the phase transitions associated with the SC order parameter of the PDW state. To capture the essential physics, we focus on the phase coherence of the Cooper pairs, assuming that amplitude fluctuations of the order parameter are frozen at low temperatures~\cite{Emery_1995, Carlson_1999, Li_2007}. Our analysis shows that the phase coherence of the SC order parameter can be effectively described by a frustrated XY model on an emergent kagome lattice, which shares the same ground-state structure and topological excitations as the original system~\cite{Rzchowski_1997, Park_2001, Harris_1992, Korshunov_2002, Andreanov_2020}. With this correspondence, the magnetoresistance oscillations with a period of $\frac{h}{6e}$ can be naturally understood as arising from the condensation of $1/3$ fractional vortex anti-vortex pairs.

Using the recently developed tensor network (TN) methods for frustrated spin systems~\cite{Vanhecke_2021, Song_2022, Song_2023a, Song_2023b}, we determine the phase structure of the antiferromagnetic (AF) XY model with nearest-neighbor (NN) interactions on kagome lattices. From the singularity of the entanglement entropy of the 1D quantum transfer operator, we find clear evidence of a single Berezinskii-Kosterlitz-Thouless (BKT) phase transition~\cite{Berezinsky_1971, Kosterlitz_1973, Kosterlitz_1974}, driven by the unbinding of fractional vortex-antivortex pairs with $1/3$ vorticity. In the low-temperature phase, we find a quasi-long-range order (quasi-LRO) in the spin variable $\exp(i3\theta)$, while the phase coherence of integer vortices is destroyed, as indicated by the exponential decay of $\exp(i\theta)$ in correlation functions. This clarified phase structure demonstrates a scenario in which conventional charge-$2e$ superconductivity arising from Cooper pairs is suppressed, and an unconventional charge-$6e$ superconducting state formed by bound states of ``Cooper sextuples'' emerges as the dominant order.

\begin{figure}[t]
\centering
\includegraphics[width=\linewidth]{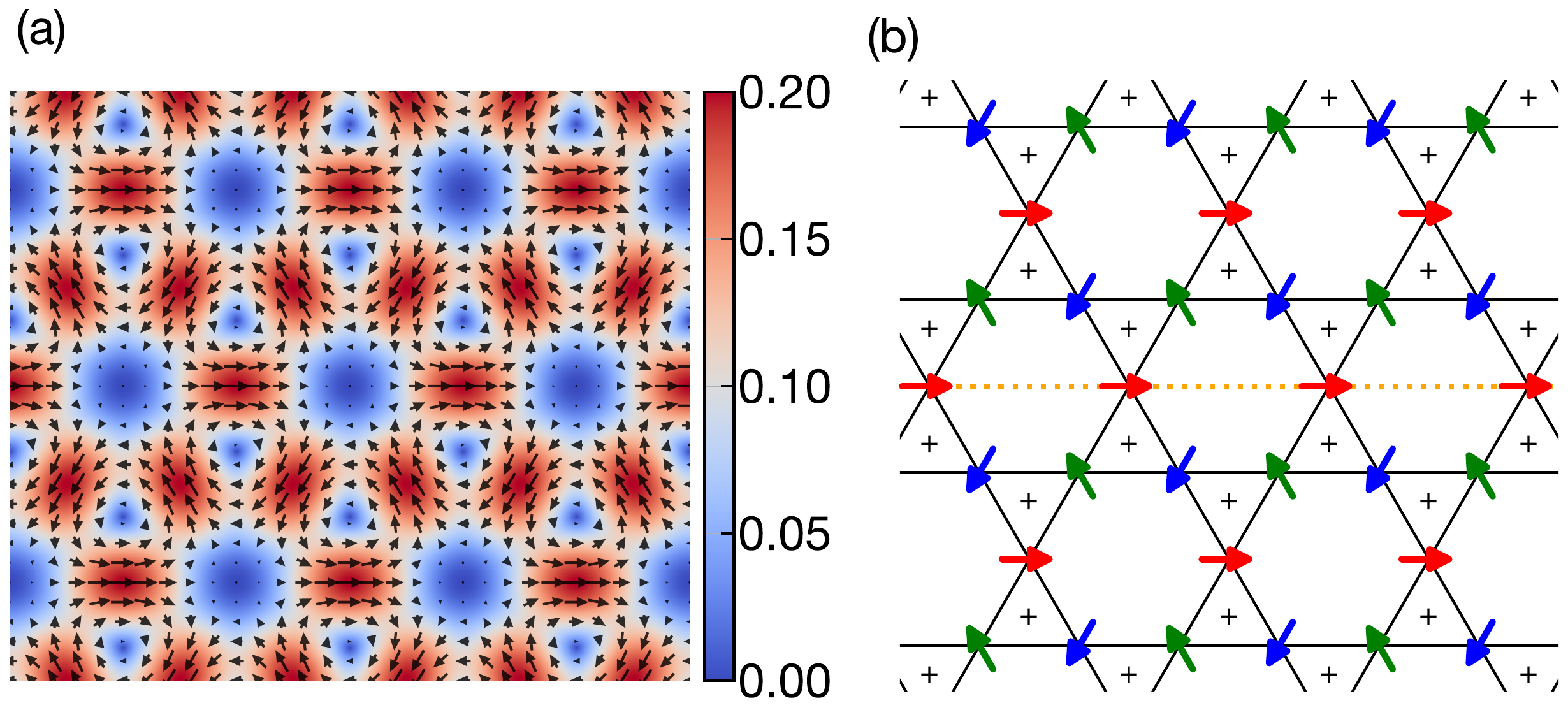}
\caption{
(a) Spatial distribution of the complex PDW order parameter, with amplitude represented by the color intensity and phase denoted by arrows.
(b) The phase distribution of the PDW state on the emergent kagome lattice represented by XY spins as  one of the ground states of the
AFXY model with chirality $\tau_i=+1$ on each triangular plaquette. The dotted orange line denotes a possible way to excite a zero energy domain wall.}
\label{fig: pdw}
\end{figure}

{\textit{Antiferromagnetic kagome XY model}}---Since the microscopic interaction is currently unknown, we consider the simplest commensurate $3\bm{Q}$ complex PDW order parameter allowed by the $P6/mmm$ space group~\cite{Agterberg_2011, Lin_2024}:
\begin{equation}
\Delta_{\text{pdw}}(\bm{r}) = \Delta e^{i\theta} \sum_{\eta=1}^{3} e^{i2(\eta-1)\pi/3} \cos(\bm{Q}_\eta \cdot \bm{r} + \phi_\eta),
\end{equation}
where $\Delta$ is the amplitude, $\theta$ is the global superconducting SC phase, $\bm{Q}_\eta$ are the modulation wave vectors and the relative phases between different modes are chosen as $\phi_\eta = 0$.

The spatial distribution of the PDW order parameter is illustrated in Fig.~\ref{fig: pdw} (a), where the amplitude is represented by the color intensity and the phase is indicated by the arrows. The amplitude reaches its minimum at the centers of the triangles and hexagons, while its maximum forms an emergent kagome lattice. Consequently, the phase coherence of the order parameter can be effectively described by an XY spin model, as shown in Fig.~\ref{fig: pdw} (b), where each spin corresponds to a phase angle at the amplitude maxima.

The phase distribution in Fig.~\ref{fig: pdw} (b) represents one of the massive degenerate ground states of the AFXY model on the kagome lattices
\begin{equation}
H=J\sum_{\langle i,j\rangle} \cos(\theta_i-\theta_j),
\label{eq: afxy}
\end{equation}
where $J>0$ and the sum runs over all nearest neighbors. Besides the overall $U(1)$ degrees of freedom, frustration within each triangular plaquette induces chiral degrees of freedom. To minimize energy, the three XY spins in a triangle must differ by an angle of $2\pi/3$, rotating either clockwise or counterclockwise. This assigns to each plaquette a chirality $\tau_i = \pm 1$. The extensive degeneracy of these chiralities can be mapped to a 3-state AF Potts model, with a residual entropy of $1.26k_B$ per site~\cite{Huse_1992}. Such degeneracy results from the excitation of zero-energy domain walls, such as the dotted orange line in Fig.~\ref{fig: pdw} (b), along with a permutation of green and blue spins in the upper plane.

The ferromagnetic ordering of the chirality in the PDW state can be stabilized by a small AF next-to-nearest-neighbor (NNN) interaction, corresponding to a charge-$2e$ SC state. The freeze of chiral degrees of freedom is due to the finite energy of the domain walls~\cite{Korshunov_2002}. However, as the temperature increases, the energy barrier is overcome, allowing domain walls to proliferate and destroy the long-range chiral order. Since the phase angles of the spins across a domain wall can change by $\pm2\pi/3$, these chirality fluctuations allow the phase angle between distant spins to vary freely by multiples of $2\pi/3$. Such fluctuations grow with distance, destroying the quasi-LRO of the XY spins of $\exp(i\theta)$, and leaving charge-$6e$ as a vestigial state at higher temperatures, where $\exp(i3\theta)$ remains algebraic under the $\pm2\pi/3$ phase fluctuations.

\begin{figure}[t]
\centering
\includegraphics[width=\linewidth]{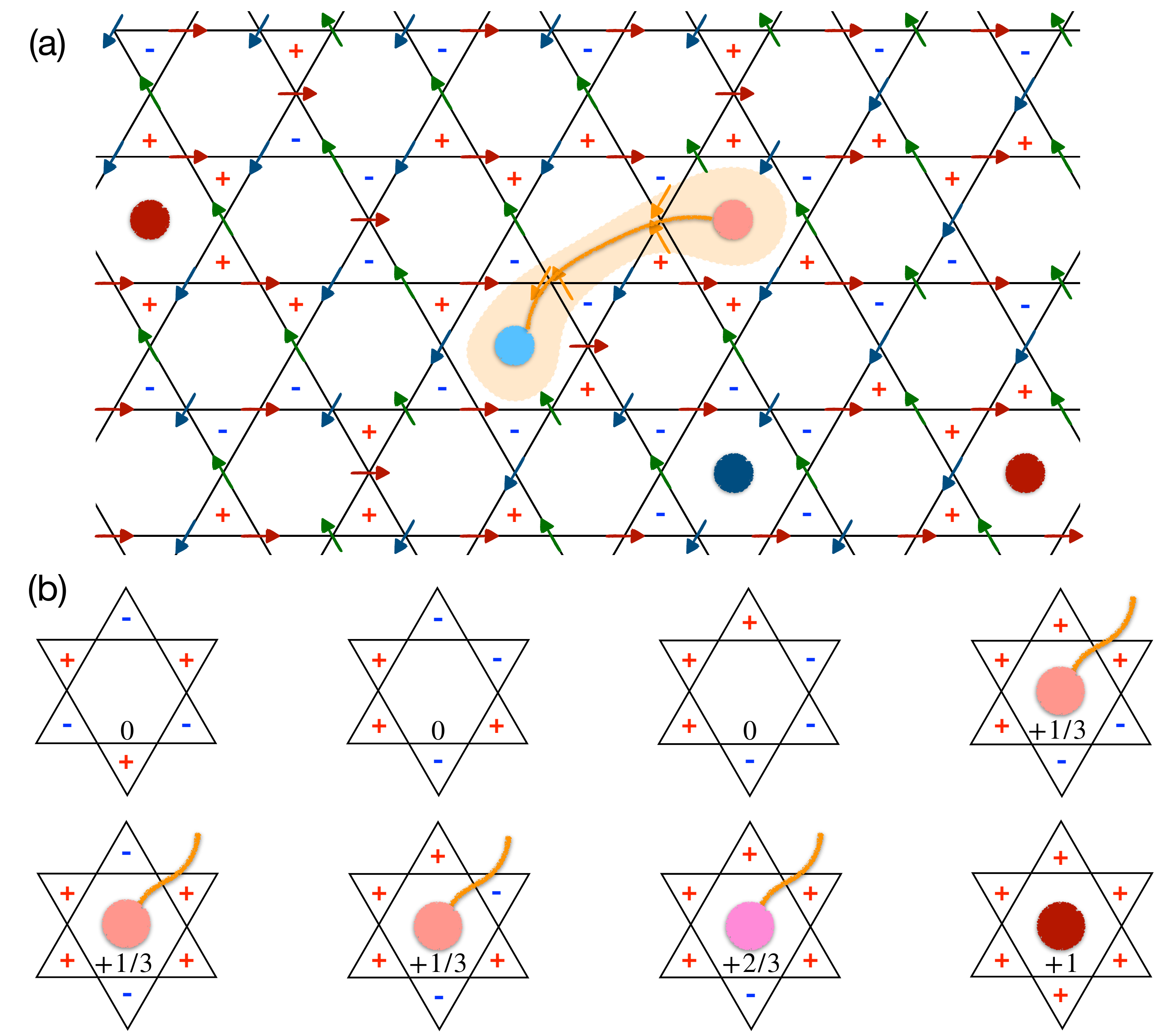}
\caption{
(a) A bound pair of $1/3$ vortices (red and blue circles) connected by a topological string (orange line) is formed by switching the chiralities of two adjacent triangles in one of the degenerate groundstate configurations. Integer vortices (dark red and blue circles) can freely excite.
(b) Topological charges of vortices corresponding to different domain wall structures. A corresponding set of negative vortices can be obtained by inverting all the signs of chirality. Phase mismatches between spins are indicated by orange lines. }
\label{fig: fvortex}
\end{figure}

{\textit{Excitations of $1/3$ fractional vortices}}---In the absence of chiral order, the concept of excess charges offers a clearer physical picture than complex domain wall configurations~\cite{Korshunov_1986}. The topological charge of a vortex at the center of a hexagon is determined by the chiralities of the six surrounding triangles $q_h = \frac{1}{3}\sum_{t=1}^6 q_t$, where $q_t = \pm \frac{1}{2}$ corresponds to chiralities $\tau_t = \pm 1$. The relationship between vortex charges and chirality configurations is shown in Fig.~\ref{fig: fvortex} (b). An alternative set of vortex configurations can be generated by inverting all chiralities on the triangles. Integer vortex excitations do not disrupt the constraint that each spin must differ from its neighbors by $2\pi/3$, as they do not introduce a phase mismatch on any triangle. In contrast, fractional vortices create a phase mismatch among the six surrounding spins of a hexagon, as shown by the orange line extending from the center of hexagons in Fig.~\ref{fig: fvortex} (b), and incur a higher energy cost.

At low temperatures, the AFXY model \eqref{eq: afxy} can be mapped to a 2D Coulomb gas with charges $Q_h = \frac{1}{3}\sum_{t=1}^6 \tau_t + m_h$, where $m_h$ are integers defined at the hexagon centers~\cite{Rzchowski_1997}. The constraint $\sum_{t=1}^6 \tau_t \equiv 0 \ (\text{mod}\ 6)$ ensures that integer vortices can freely excite at low temperatures, whereas fractional vortices are bound in pairs due to their higher energy cost. As shown in Fig.~\ref{fig: fvortex} (a), a pair of $\pm \frac{1}{3}$ vortices can be created by switching the chiralities of two adjacent triangles. The fractional vortex pairs are connected by a topological string, along which a phase mismatch of $2\pi/3$ exists. Upon increasing the temperature, these fractional vortices unbind at a BKT transition, marking the destruction of quasi-LRO in $\exp(i3\theta)$.

The flux quantization of $\frac{h}{6e}$ arises from the quasi-long-range order (QLRO) of $\exp(i3\theta)$, where the SC order parameter is $\Delta_{6e} \propto \Delta_{Q_a} \Delta_{Q_b} \Delta_{Q_c} = \Delta^3 \exp(i3\theta)$, describing a six-electron bound state~\cite{Agterberg_2011, Zhou_2022}. From the minimal coupling relation $\nabla \theta = \frac{6e}{\hbar} \vec{A}$, integrating around a closed loop enclosing a vortex core gives the flux quantization condition $\Phi_{6e} = \frac{h}{6e}$, corresponding to the $1/3$ vortex in the phase field.

\begin{figure}[t]
\centering
\includegraphics[width=\linewidth]{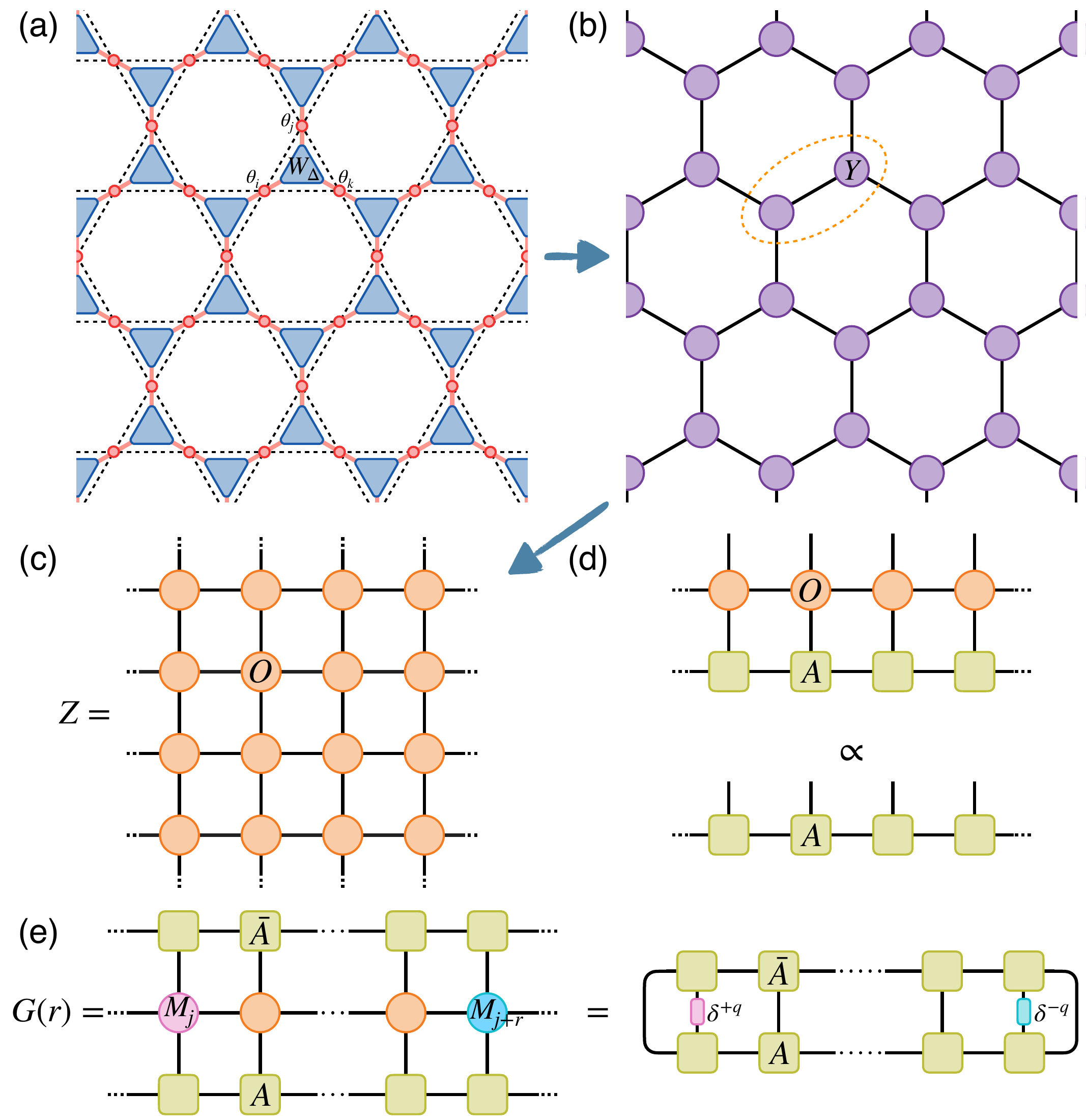}
\caption{
(a) Tensor network with continuous indices. $W$ tensors represent Boltzmann weights on triangles, and the red dot integrates joint $\theta$ variables between two triangles.
(b) Tensor network with discrete indices obtained via Fourier transformations.
(c) Uniform infinite tensor network represented by the local tensor $O$ by combining two neighboring $Y$ tensors.
(d) Fixed-point MPS eigenequation for the 1D transfer operator.
(e) Two-point correlation function calculated by contracting a sequence of channel operators with canonical MPS.}
\label{fig: tensor}
\end{figure}

{\textit{Tensor network methods}}---The TN approach is particularly well-suited for investigating the low-temperature regime of frustrated models, as the massive degeneracy induced by strong frustration can be efficiently encoded in the local tensors~\cite{Vanhecke_2021, Song_2022, Song_2023a, Song_2023b}. The first step in applying the TN method is to express the partition function of the lattice model \eqref{eq: afxy} as a TN. To account for the chiral degrees of freedom associated with each triangle, the partition function is decomposed into a product of local Boltzmann weights as shown in Fig~\ref{fig: tensor} (a)
\begin{equation}
Z=\prod_i\int\frac{d\theta_i}{2\pi}\prod_{\Delta}W_{\Delta},
\end{equation}
where $W_{\Delta}(\theta_i,\theta_j,\theta_k)=e^{-\beta H_{\Delta}}$ is a three-leg tensor with continuous $U(1)$ indices depicted by red dots on the lattice sites. And the local Hamiltonian $H_{\Delta}(\theta_i,\theta_j,\theta_k)$ on each triangle is defined as $H_{\Delta}=J\sum_{\langle i,j\rangle\in{\Delta}}\cos(\theta_i-\theta_j)$.

Then a Fourier transformation is applied to bring the local tensor $W_{\Delta}$ onto a discrete basis, as displayed in Fig~\ref{fig: tensor} (b),
\begin{equation}
Y_{n_{1},n_{2},n_{3}} =\prod_{i=1}^{3}\int \frac{d\theta _{i}}{2\pi }W_{\Delta}(\theta _{1},\theta _{2},\theta _{3}) e^{in_1\theta_1+in_2\theta_2+in_3\theta_3}.
\end{equation}
From a saddle point approximation, we can get the asymptotic formula of $Y_{n_{1},n_{2},n_{3}}\propto\exp[-\frac{1}{3\beta}(n_1^2+n_2^2+n_3^2)]$, implying an exponential decay with tensor indices $n$. Hence, we can safely truncate the local $Y$ tensors with finite bond dimensions. Finally, the translation-invariant local tensor $O$ is achieved by combining a pair of $Y$ tensors $O_{n_1,n_2}^{n_3, n_4}=\sum_{n_5}Y_{n_{1},n_{2},n_{5}}Y_{n_{3},n_{4},n_{5}}$. And the uniform TN representation of the partition function displayed in Fig~\ref{fig: tensor} (c) is given by
\begin{equation}
Z=\mathrm{tTr}\prod_s O_{n_1,n_2}^{n_3, n_4}(s),
\end{equation}
where “tTr” denotes the tensor contraction over all auxiliary bonds.

In the thermodynamic limit, the fundamental object for the calculation of the partition function is the row-to-row transfer matrix $T(\beta)=\mathrm{tTr}\left[\cdots O(p) O(q) O(r) \cdots\right]$, which is analogous to the matrix product operator (MPO) in a 1D quantum chain. The contraction of the tensor network reduces to finding the leading eigenvalue and eigenvectors
\begin{equation}
T(\beta)|\Psi(A)\rangle=\Lambda_{\max}|\Psi(A)\rangle,
\label{eq: eigeneq}
\end{equation}
as shown in Fig~\ref{fig: tensor} (d), where $|\Psi(A)\rangle$ is the leading eigenvector represented by matrix product states (MPS) made up of uniform local A tensors whose auxiliary bond dimension $D$ controls the accuracy of the approximation.~\cite{Zauner_2018}. The fixed-point eigenequation can be accurately solved by the variational uniform matrix product state (VUMPS) algorithm~\cite{Zauner_2018, Vanderstraeten_2019, Nietner_2020}.

Once the fixed points are achieved, various physical quantities can be efficiently calculated within the TN framework. The entanglement entropy of the 1D quantum correspondence is directly obtained from the Schmidt decomposition of $|\Psi(A)\rangle$ as $S_E = -\sum_{\alpha=1}^{D} s_{\alpha}^2 \ln s_{\alpha}^2$, where $s_{\alpha}$ are the singular values. Local observables are computed by inserting the corresponding impurity tensors into the tensor network for the partition function. For $e^{iq\theta_j}$, the impurity tensor modifies the $U(1)$ charge conservation law $\delta_{n_1+n_2+n_3,0}$ in the $Y$ tensor to $\delta_{n_1+n_2+n_3+q,0} = \sum_{n_4} \delta_{n_1+n_2+n_4,0} \delta_{n_3+q,n_4}$, introducing an additional unbalanced delta tensor $\delta^{+q} = \delta_{n_1+q,n_2}$.

Using the MPS fixed point, the contraction of the tensor network containing impurity tensors is reduced to the trace of an infinite sequence of channel operators, which can be further simplified into the contraction of a smaller network. As shown in Fig.~\ref{fig: tensor} (e), the two-point correlation function is given by
\begin{equation}
G_q(r) = \langle \cos(q\theta_i - q\theta_{i+r}) \rangle = \langle e^{iq\theta_i} e^{-iq\theta_{i+r}} \rangle,
\label{eq: clfn}
\end{equation}
where the canonical form of the MPS is employed to ensure efficient contraction.

\begin{figure}[t]
\centering
\includegraphics[width=\linewidth]{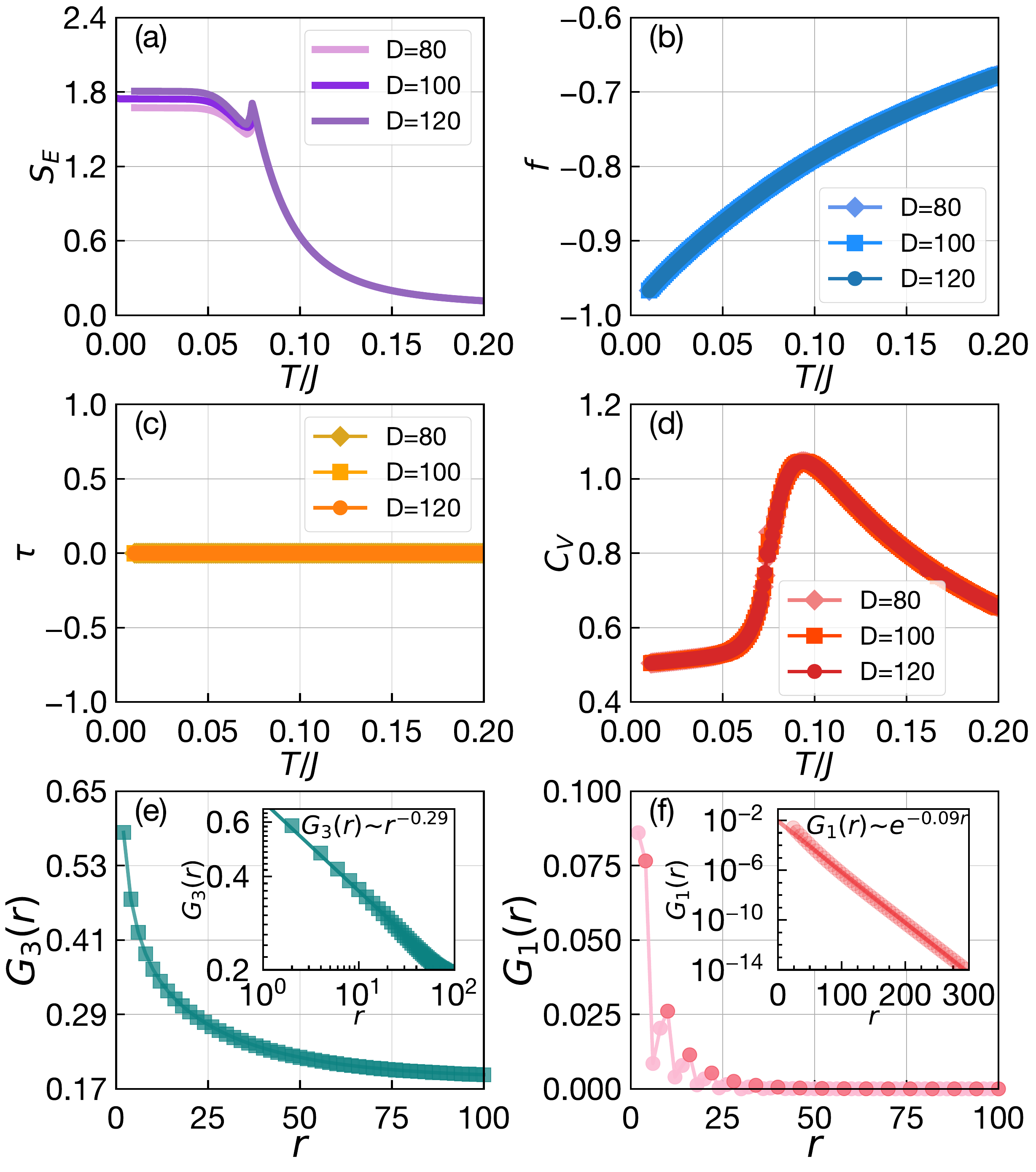}
\caption{
(a) The entanglement entropy $S_E$ as a function of temperature for different MPS bond dimensions. For $D = 100$, calculations extend to the lowest temperature of $T/J = 10^{-5}$.
(b) The free energy remains smooth across all temperatures, indicating a continuous phase transition.
(c) The chirality $\tau$ is zero at all temperatures, confirming the absence of long-range chiral order.
(d) The specific heat $C_V$ exhibits a rounded bump above the critical temperature $T_c$, characteristic of a BKT transition.
(e) The correlation function $G_3(r)$ for $1/3$ fractional vortices shows power-law decay at $T/J = 0.07$ below $T_c$.
(f) The correlation function $G_1(r)$ for integer vortices decays exponentially at $T/J = 0.07$.}
\label{fig: result}
\end{figure}

{\textit{Thermodynamic quantites}}---The entanglement entropy of the fixed-point MPS for the 1D quantum correspondence exhibits singularities at the critical temperatures, providing a clear criterion for phase transitions in the thermodynamic limit. As shown in Fig.~\ref{fig: result} (a), $S_E$ displays a sharp singularity at $T/J \simeq 0.075$, indicating a single phase transition at a rather low temperature. The peak position remains almost unchanged for MPS bond dimensions ranging from $D = 80$ to $120$, allowing precise determination of the transition temperature, which agrees well with theoretical predictions of $T_c/J\approx \frac{\pi/\sqrt{3}}{72}$ for the unbinding temperature of $1/3$ vortex pairs~\cite{Korshunov_2002}. The TN approach enables efficient exploration of the extremely low-temperature regime previously inaccessible. For $D = 100$, we calculate $S_E$ at temperatures as low as $T/J = 10^{-5}$ and find that it remains smooth, ruling out the possibility of a transition to long-range chiral order at low temperatures.

We further investigate the thermodynamic properties to gain insight into the phase transition. The thermodynamic quantities converge well and remain consistent across different MPS bond dimensions. The free energy per site is calculated from the eigenvalues of \eqref{eq: eigeneq} as $f = -\frac{1}{3\beta} \ln \lambda$. As shown in Fig.~\ref{fig: result} (b), the free energy is perfectly smooth, indicating a continuous phase transition. The expectation value of chirality is given by $\tau = \frac{2}{3\sqrt{3}} \sum_{\langle i,j \rangle} \langle \sin(\theta_i - \theta_j) \rangle$. As shown in Fig.~\ref{fig: result} (c), $\tau$ equals zero at all temperatures, confirming the absence of long-range chiral order due to strong chirality fluctuations. Meanwhile, we observe $\langle \cos(\theta_i - \theta_j) \rangle \to -1/2$ at low temperatures, implying that the angle between NN spins approaches $2\pi/3$. Additionally, the specific heat, derived as $C_V = \frac{du}{dT}$, shows a rounded bump around $T/J = 0.095$ as displayed in Fig.~\ref{fig: result} (d), higher than the transition temperature determined from the entanglement entropy. The smooth behavior of thermodynamic properties rules out the possibility of the first- or second-order transitions from the lifting of chirality degeneracy, which is consistent with a BKT transition.

{\textit{Correlation functions}}---To investigate the nature of the phase transition, we calculate two correlation functions, $G_1(r)$ and $G_3(r)$, for integer and fractional vortices, as defined in \eqref{eq: clfn}.  A comparison between the two correlation functions in the low-temperature phase is shown in Fig.~\ref{fig: result} (e) and (f). At $T/J = 0.07$ below $T_c$, $G_3(r)$ exhibits power-law decay with distance, indicating the quasi-LRO of fractional vortices, while $G_1(r)$ decays exponentially, reflecting the absence of LRO for integer vortices. The exponential decay in $e^{i\theta}$ can be attributed to the free fluctuations among different ground-state chirality patterns, where the uncertainty in the phase difference in multiples of $2\pi/3$ grows with distance and destroys the long-range correlations in $\theta$ fields. Interestingly, $G_1(r)$ shows damped oscillations at short range but decays rapidly at larger distances. These oscillations arise because spin-wave fluctuations promote short-range chirality ordering but are too weak to lift the ground-state degeneracy. Above $T_c$, both $G_1(r)$ and $G_3(r)$ decay exponentially, signaling that the system transitions into the disordered phase.

This distinction underscores the contrasting roles of integer and fractional vortices. At low temperatures, the proliferation of integer vortices disrupts phase coherence between charge-$2e$ Cooper pairs. Meanwhile, $1/3$ vortex pairs exhibit quasi-LRO below $T_c$, reflecting the condensation of Cooper sextuples—a hallmark of charge-$6e$ superconductivity~\cite{Ge_2024}. Above $T_c$, the system transitions to the normal phase through a BKT transition driven by the dissociation of fractional vortices.

{\textit{Discussion and conclusion}---In this work, we investigated the unconventional superconductivity in the kagome SC $\mathrm{CsV_3Sb_5}$~\cite{Ge_2024}, focusing on the emergence of charge-$6e$ SC at higher temperatures above the conventional charge-$2e$ SC state. By modeling the phase coherence of the SC order parameter with a frustrated XY model with NN interactions on an emergent kagome lattice, we demonstrated that the condensation of fractional vortices with $1/3$ vorticity plays a crucial role in stabilizing the charge-$6e$ SC. Using state-of-art TN methods~\cite{Vanhecke_2021, Song_2022, Song_2023a, Song_2023b}, we clarified the nature of the phase transition. The critical temperature is precisely determined from the entanglement entropy singularity in the 1D quantum correspondence. A systematic analysis of thermodynamic properties and correlation functions in the thermodynamic limit reveals a single BKT transition at $T_c/J \simeq 0.075$, driven by the unbinding of $1/3$ fractional vortex-antivortex pairs. Below $T_c$, the correlation functions for fractional vortices exhibit power-law decay, while those for integer vortices decay exponentially. The absence of chiral long-range order or phase coherence between integer vortices demonstrates that the low-temperature phase is dominated by charge-$6e$ SC, with no contribution from charge-$2e$ SC.

We would also like to point out that the phase transition from charge-$2e$ SC to charge-$6e$ SC is not addressed in this work. The stabilization of the PDW state can be achieved by further incorporating NNN interactions into the AFXY kagome model, where the destruction of the PDW order is expected to occur via a first-order phase transition~\cite{Korshunov_2002, Kakizawa_2024, Lin_2024}. However, the full phase diagram, including both NN and NNN interactions, remains an endeavor for future research. Significant challenges arise in its exploration, such as the large bond dimensions required for TN methods and the excessively long relaxation times encountered in Monte Carlo simulations. A systematic study of the complete phase diagram would offer valuable insights into the interplay of interactions and frustration in these systems.

Alternatively, the PDW order could be destabilized by weak disorder or through mechanisms of 2D thermal melting~\cite{Nelson_1978, Nelson_1979, Young_1979, Agterberg_2008, Agterberg_2011, Agterberg_2020}. Further investigation of these possibilities may shed light on unresolved issues, including the origin of the charge-$4e$ oscillations observed in smaller samples~\cite{Ge_2024}. These phenomena, which are not addressed in the current study, highlight the need for future theoretical and experimental efforts to more fully understand the rich physics of kagome superconductors.

\begin{acknowledgments}
{\textit{Acknowledgments}}---The authors are very grateful to Yu-Tong Lin for stimulating discussions. The research is supported by the National Key Research and Development Program of China (Grant No. 2023YFA1406400).
\end{acknowledgments}


\end{document}